\RequirePackage{fix-cm}
\documentclass[smallextended]{svjour3}       
\smartqed  
\usepackage{graphicx}
\usepackage{subcaption}
\usepackage{float}
\usepackage{multirow}
\usepackage{amsmath,amssymb,amsfonts}
\usepackage{mathtools, cuted}
\usepackage{hyperref}

\begin{document}

\title{Distributed Application Provisioning over Ethereum based private and permissioned Blockchain: Availability modeling, capacity, and costs planning
}


\author{Carlos Melo, Jamilson Dantas, Paulo Pereira and Paulo Maciel
}


\institute{Carlos Melo, Jamilson Dantas, Paulo Pereira, and Paulo Maciel \at
              Centro de Informática, Universidade Federal de Pernambuco, Recife, Brazil \\
              Tel.: +55-87-999265912\\
              \email{{casm3,jrd,prps,prrm}@cin.ufpe.br}}           

\date{Received: date / Accepted: date}

\maketitle

\begin{abstract}
Blockchain and Cloud Computing are two of the main topics related to the distributed computing paradigm, and in the last decade, they have seen exponential growth in their adoption. Cloud computing has long been established as the main mechanism to test, develop, and deliver new applications and services in a distributed manner across the World Wide Web. Large data centers host many services and store petabytes of user data.  Infrastructure and services owners rule the access to data and may even be able to change contents and attest to its veracity. Blockchain is a step towards a future where the user's data are considered safer, besides being public. Advances in blockchain-based technologies, now, support service provisioning over permissioned and private infrastructures. Therefore, organizations or groups of individuals may share information, service even if they do not trust each other, besides supporting infrastructure management tasks. This paper presents and evaluates models for assessing the availability and capacity-oriented availability of cloud computing infrastructures. It aims at running Blockchain's distributed applications based on the Ethereum blockchain platform and the required expenses to perform service delivery in public and private infrastructures. Most of the obtained results also apply to other blockchains based platforms. 
\end{abstract}

\section{Introduction}
\label{intro}
Most of the blockchain-based applications try to solve trust problems related to data's privacy, ownership, and control, which are often in possession of a single entity or a group of entities that run massive data centers or that own and manage a respective service, whether this is a social network or even a banking system \cite{sukhwani2017performance}. Usually, those entities may modify and certify the authenticity of the altered data; also, they can charge high fees for the carried transactions, which goes against what is proposed by blockchain platforms \cite{guy2015}.

Besides Blockchain, cloud computing has emerged as one of the leading technologies associated with the distributed computing paradigm, being a direct result of the virtualization technology evolution \cite{vaqueiro2009}. Due to virtualization, data centers possess a mechanism to test, refine and develop applications by reducing idle resources, increasing the profit margin, and optimizing resource sharing by hosting and providing any service (infrastructure, software, and platform) over the Internet \cite{mell2011nist}.

Anyone may contract and offer services over cloud computing environments, this on-demand delivery of IT resources can be done via the Internet with pay-as-you-go pricing mechanism, always concerning a Service Level Agreement (SLA), where acceptable rates of availability and performance are established between both environment provider and client or application provider\cite{amazon2020}. Besides contracting the cloud computing environments from well-known providers, such as Google Cloud, Amazon AWS and Microsoft Azure, one could deploy their private infrastructure, improving the security and control over the environment, reducing expenses by doing it with free and open-source software, such as Docker and OpenStack.

Blockchain and cloud computing will promote the concept of distributed cloud, which is characterized by adding new values and responsibilities to service users \cite{onik2019}. Therefore, users should be able to use and host distributed services, without the need for large data centers or the information control by large institutions, as well as deploying their own private infrastructure. However, a transition period is required to identify the feasibility of this paradigm, at first we need at least a platform that can be executed either on computer servers or mobile devices, as well as an established pattern to run distributed applications. 

The Ethereum platform improved the support for the rapid development of distributed applications by providing a native programming language (Solidity) \cite{buterin2013ethereum}. Docker container engine has become a community standard for distributed services provision across large and distributed data centers \cite{hightower2017kubernetes}. 

Ethereum and Docker are stable technologies; each one is present in the daily lives of many persons around the world. However, combining both blockchain platforms and virtualization technology is still in its early steps. Despite that, works have been done to evaluate essential metrics associated with service provisioning, such as the system's availability \cite{aniello2017,roehrs2019}, performance \cite{sukhwani2017performance,pongnumkul2017performance} and capacity-oriented availability \cite{melo2017capacity,torquato2175models}. Most of them focus only on cloud computing architectures, services, and distributed applications, other ones focused on blockchain platforms. However, as for the best of our knowledge, the present paper is the first one to evaluate the availability and assess the impact of the failures and repairs routines over the capacity planning of blockchain platforms.

As the contribution of this paper, we can list:
\begin{itemize}
\item Availability models for blockchain provisioning over cloud computing and virtualized environments managed by Docker;
\item Sensitivity analysis evaluation to detect bottlenecks in service provisioning;
\item Capacity-oriented availability planning for service provisioning over architectures with a large amount of blockchain computing nodes;
\item Cost evaluation and comparison between public and private cloud computing environments on blockchain-as-a-service provisioning.
\end{itemize}

The remained of this paper is organized as follows. Section \ref{sec:rltd} presents the works that underlie this research. Already Section \ref{sec:background} shows the background. While the Section \ref{sec:methodology} presents the support methodology that enable the proposed models construction, as presented in Section \ref{sec:architectures}. Section \ref{sec:case} provides the studies used to demonstrate how feasible are the proposed models. Finally, Section \ref{sec:conclusions} presents the final remarks and future works.

\section{Related Works}
\label{sec:rltd}
Over the last few years, authors have devoted their efforts to study and evaluate performance and dependability attributes related to blockchain and cloud computing technologies, in this section, we present some of these remarkable works. To the best of our knowledge, these are the closest related works to what is proposed by the current paper.

Previously in Melo et al., \cite{melo2018dependability,melo2019}, we have evaluated a blockchain-as-a-service infrastructure considering Dynamical Reliability Block Diagrams (DRBD) and a set of architectures containing the minimum requirements to deploy a Hyperledger platform over a virtualized environment managed by docker and OpenStack cloud computing platform. The current work is an extension of these two, and are the closest to what is proposed here, which evaluates the availability, capacity-oriented availability, and deployment expenses of an Ethereum private environment. Also, the proposed models here are hierarchically distributed, both Reliability Block Diagrams (RBDs) and Stochastic Petri Nets (SPN) are used in order to represent priority repairs between the nodes on Ethereum environment, we have also used analytical models provided from RBD's K-out-of-N (KooN) blocks to calculate the capacity-oriented availability \cite{paulo2011,MartinsMaciel2016}.

Pongnumkul et al. 2017 \cite{pongnumkul2017performance} proposed a performance evaluation methodology for blockchain environments, and evaluated the Ethereum and Hyperledger Fabric platforms. The authors concluded that the Hyperledger Fabric has a large throughput and lower latency than Ethereum, which is a clear result of the consensus protocol. Ethereum uses Proof-of-Work (PoW), while the Hyperledger Fabric worked with the Practical Byzantine Fault Tolerance (PBFT) protocol, but now the current versions of the Hyperledger Fabric platform work with Kafka.

Zhang et al., 2018 \cite{zhang2018method} proposed a method to predict the performance of blockchain environments based on analytical modeling; they used the Ethereum platform to demonstrate the feasibility of the proposed methodology. The associated metrics evaluated were transactions per second, and the storage space required to store the performed transactions. 

In Sukhwani et al., 2017  \cite{sukhwani2017performance}, the authors investigated whether a consensus process using the PBFT mechanism could be a performance bottleneck through Stochastic Reward Nets (SRN) modeling to compute the meantime to complete the consensus process. The PBFT protocol is one of the most used by blockchain platforms. The focus of the current paper is on the availability evaluation of the Ethereum platform. 

Later, in Sukhwani et al., 2018 \cite{sukhwani2018performance}, the authors evaluated the performance of Hyperledger Fabric through SRN models, which means that this works the entire platform performance, not only the consensus protocol. Once again, the proposed work differs from ours in the platform and the evaluated metrics. 

In Weber et al., 2017 \cite{weber2017}, the authors identified some availability limitations of two blockchains, Ethereum and Bitcoin, by measuring and gathering public data. We emphasized the evaluation of an Ethereum private blockchain availability, as well as the current acquisition and maintenance expenses, and the capacity-oriented availability, by proposing and evaluating the availability and capacity-oriented availability models. 

Roehrs et al. 2019 \cite{roehrs2019}, the authors proposed the omniPHR blockchain platform, focused on securely sharing personal health information. The authors evaluated through mathematical models and measurement the availability and performance of the proposed platform, considering besides the overall system's availability a set of performance metrics such as transactions per second, CPU utilization, memory, and network utilization. 

Aniello et al. 2017 \cite{aniello2017} have also proposed a new blockchain platform and evaluated scalability, availability, security, and performance metrics such as transactions per second and response time by considering this platform as an alternative for the traditional distributed database.

In Li et al., 2018 \cite{li2018blockchain}, the authors proposed a markovian process to express the relationship between the transaction arrival rate and the server capacity. While in \cite{torquato2175models} the authors proposed models to evaluated capacity-oriented availability (COA) of general cloud computing environments by considering Virtual Machines as their resources, considering failure and repair routines and their impact over the metric. 

As could be seen from previously presented studies, most of the work accomplished focused on performance metrics. At the same time, in the proposed paper, we consider availability models and COA's evaluation, as well as the expenses related to service provisioning in both public and private environments, considering docker containers and compute nodes as resources and performing a sensitivity analysis to detect bottlenecks and the components with the highest impact over the availability metric.
\section{Background}
\label{sec:background}
This section presents the fundamental concepts about availability evaluation, system modeling, blockchain provisioning, and the Docker engine, which is a community pattern to provide containers.
\subsection{Availability Evaluation}
\label{subsec:availability}
Dependability is the ability of a computer system to deliver a service that can be justifiably trusted \cite{Avizienis04basicconcepts}. This definition considers user perception and system behavior. Evaluating the dependability of a system is usually the result of evaluating at least one of its five attributes: reliability, availability, maintainability, integrity, and safety \cite{avizienis2001}.

\begin{figure}[ht]
\centering
\includegraphics[width=.25\textwidth]{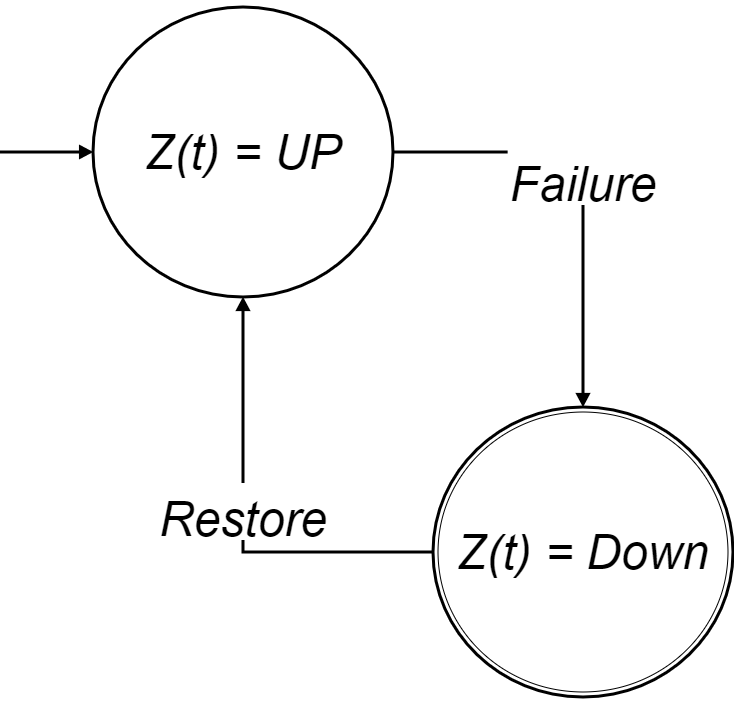}
\caption{System's States}
\label{fig:systemstate}
\end{figure}

The system's instantaneous availability, is the probability that the system is operational at a time \textit{t}. That is, $A(t) = P\left \{ Z(t) \right \} = E\left \{Z(t)  \right \}, t \geq 0$, where $Z(t) = 1$ when the system is operational, and $Z(t) = 0$, otherwise, as shown in Figure \ref{fig:systemstate}.

The Steady-state availability,  which is the one evaluated by this paper and is also called the long-run availability, is the limit of the availability function as time tends to infinity (Equation \ref{limiteav}). 

\begin{equation}
\label{limiteav}
A = \lim_{t \to \infty } A(t), t \geq 0
\end{equation}

Also, the system's availability may be represented by a ratio between the Mean Time To Failure (MTTF) and Mean Time To Repair (MTTR) of the system (Equation \ref{avmttfmttr}). 

\begin{equation}
\label{avmttfmttr}
    \text{A} = \frac{\text{MTTF}}{\text{MTTF+MTTR}}
\end{equation}

The MTTF for a system may be computed from Equation \ref{mttf}, where \textbf{R(t)} is the reliability of that system as a function of elapsed time. The Equation \ref{mttr} provides a way of computing the MTTR from the values of MTTF, availability (A), and unavailability (UA = $1-A$).  

\begin{equation}
\label{mttf}
\text{MTTF} = \int_{0}^{\infty } R(t)\partial t
\end{equation} 

\begin{equation}
\label{mttr}
\text{MTTR} = \text{MTTF} \times (\frac{UA}{A}),
\end{equation}

As expected, the system's unavailability is the counterpart of the system's availability and can be calculated by $1 - A$. Already the system's downtime, which corresponds to the period where the system is not available, is usually given in units of time, such as in hours per year, in a way that relates time and probability: $Unavailability \times 8760 h$.

The availability can also be represented by the number of nines \cite{ohmann2014achieving}, as shown in Table~\ref{tab:number_nines}. For example, a system with four 9's of availability is classified as fault-tolerant, meaning an annual downtime of nearly 1 hour.

    \begin{table}[H]
\footnotesize
    \centering
    \caption{Service availability in number of nines.}
    \begin{tabular}{p{1cm} p{1cm} p{1cm} p{1cm}}
    \hline
     \multicolumn{1}{l}{\textbf{\vtop{\hbox{\strut \# of}\hbox{\strut 9's}}}} & \multicolumn{1}{l}{\textbf{\vtop{\hbox{\strut Avail.}\hbox{\strut (\%)}}}} & \multicolumn{1}{l}{\textbf{\vtop{\hbox{\strut System}\hbox{\strut Type}}}} & \multicolumn{1}{l}{\textbf{\vtop{\hbox{\strut Downtime}\hbox{\strut (year)}}}} \\

    \hline

     \multicolumn{1}{l}{\textbf{1}} & \multicolumn{1}{l}{90} & \multicolumn{1}{l}{unmanaged} & \multicolumn{1}{l}{5 weeks} \\ 
     
     \multicolumn{1}{l}{\textbf{2}} & \multicolumn{1}{l}{99} & \multicolumn{1}{l}{managed}  & \multicolumn{1}{l}{4 days} \\ 
    
     \multicolumn{1}{l}{\textbf{3}} & \multicolumn{1}{l}{99.9} & \multicolumn{1}{l}{well managed}  & \multicolumn{1}{l}{9 hours} \\
     
     \multicolumn{1}{l}{\textbf{4}} & \multicolumn{1}{l}{99.99} & \multicolumn{1}{l}{fault tolerant}  & \multicolumn{1}{l}{1 hour} \\
     
     \multicolumn{1}{l}{\textbf{5}} & \multicolumn{1}{l}{99.999} & \multicolumn{1}{l}{high availability}  & \multicolumn{1}{l}{5 minutes} \\
     
     \multicolumn{1}{l}{\textbf{6}} & \multicolumn{1}{l}{99.9999} & \multicolumn{1}{l}{very high availability}  & \multicolumn{1}{l}{30 seconds} \\
     
     \multicolumn{1}{l}{\textbf{7}} & \multicolumn{1}{l}{99.99999} & \multicolumn{1}{l}{ultra availability}  & \multicolumn{1}{l}{3 seconds} \\

    \hline

    \end{tabular}
    
      \label{tab:number_nines}
\end{table}

The system's availability of associated metrics and values usually can be evaluated through simulation, measurement, and analytical models. The latter was chosen due to the high-level system view provided, as well as their great flexibility in adapting parameters and achieving results faster compared to the other two evaluation techniques. Regarding analytical modeling, two main trends are highlighted:

\begin{itemize}
    \item \textbf{Combinatorial or non-state-space} models capture conditions that make a system fail (or to be working) regarding structural relationships between the system's components \cite{Trivedi1996};
    \item Already \textbf{State-space models} represent the system's behavior (failures and repair activities) by a set of states and the occurrence of events that can be expressed as rates or distribution functions. These models allow representing more complex relationships between the system's components than combinatorial models do \cite{matos2015}.
\end{itemize}

Reliability Block Diagram (RBD) \cite{paulo2011} and Fault Tree (FT) \cite{Malhora1994} are among the most prominent combinatorial modeling formalisms, whereas Petri Nets (PN), Continuous Time Markov Chain (CTMC) and Stochastic Automata Networks are well-known state-space modeling formalisms \cite{paulo2011,Garg1995,Trivedi1996}. In this paper, RBDs are used to describe the system part that does not present any dependency between components.

Petri Nets \cite{murata1989} is a family of formalism well suited for modeling several system types, including concurrency, synchronization, communication mechanisms, besides supporting deterministic and probabilistic delays. This paper adopts a particular extension of Petri nets, namely Stochastic Petri Nets (SPN) \cite{molloy1981}, which allows the association of stochastic delays to timed transitions using the exponential distribution. We present an SPN that is used to estimate the overall system's availability associated with a set of metrics such as the system's annual downtime for the blockchain-based application provisioning, this SPN receive input values extracted from an RBD and is used as an input for another RBD, used to estimate the availability of an extensive infrastructure as well as the capacity-oriented availability (COA).

\subsection{Docker, Blockchain and Ethereum Platform}
\label{subsec:nuvem}

Docker is a set of related technologies that includes a container image format and a runtime library, which manages containers' life cycle through APIs and a command-line tool \cite{arundel2019cloud}. Docker is the actual pattern for container creation and management; this technology enables users and companies to virtualize their resources and the application only. The virtualized application uses the Kernel of the host operating system, which makes the container images a lot smaller than those from virtual machines managed by hypervisors and require an entire operating system virtualization to run the provided service as KVM and VirtualBox does. This paper evaluates the use of Docker containers running Ethereum blockchain images \footnote{Ethereum: https://hub.docker.com/u/ethereum} to perform a distributed service.

Blockchain technology was first introduced by Satoshi Nakamoto, back in 2008 \cite{nakamoto2008bitcoin}, on the same paper, what later would become just an application of Blockchain was entitled as Bitcoin, an electronic and monetary payment system capable of preventing double expend, as well as a decentralized application, where computational power was the key to keeping data away from malicious users. In 2013, Vitalik Butterin proposed the Ethereum blockchain \cite{buterin2013ethereum}, the first blockchain platform to include a complete program language known as Solidity, and the possibility to develop a distributed application (Dapp) that could be maintained by both users and service providers. The Ethereum platform was the first platform that changes the way that we see blockchain technologies; today, Blockchain is not only a payment system but a distributed platform ready to host any kind of distributed service.

The Ethereum platform is portable, such as the Java computing Language, where the developed applications may run in any environment due to the Java Virtual Machine (JVM). The Ethereum distributed applications (Dapps), which are written in Solidity, runs over the Ethereum Virtual Machine (EVM).


Already an Ethereum node is the client that enables one to participate in an Ethereum network, and many Ethereum clients can turn a device into an Ethereum node. Some of the clients are provided by the Ethereum community, such as Geth and Trinity, while others are from third-party developers, which is the case of Parity.

The Ethereum platform can be provided as a cloud service. Blockchain-as-a-Service became a trend among those service providers and will be even more present in daily lives since banking services, and other distributed applications will migrate from traditional cloud models to a more safe and secure mechanism based on Blockchain.

\begin{figure}[htpb]
    \centering
    \includegraphics[width=.7\textwidth]{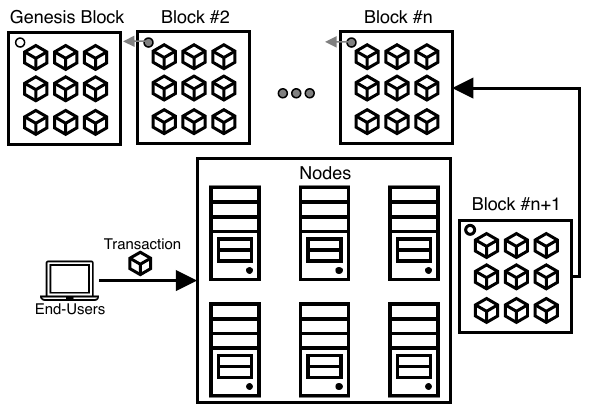}
    \caption{How does Blockchain Works?}
    \label{fig:blockchain}
\end{figure}

By general means, a blockchain application is based on shared ledgers used to store transactions' records into entities called blocks, inserted one after another into a kind of linked list \cite{dummies}. Each block, besides storing data about the performed transactions, also has a pointer to the immediately preceding block. Figure \ref{fig:blockchain} shows how a typical blockchain works.

End-user requests to a peer-to-peer network composed primarily of conventional machines or dedicated servers, these machines receive the names of nodes in a P2P architecture and are responsible for executing, ordering, and validating the required transactions. Each transaction is inserted into a block with other transactions, and each block is inserted into what we call a blockchain.
\subsection{Sensitivity Analysis}
Sensitivity analysis is adopted to determine the most influential factors over a metric of interest \cite{frank1978}. In this paper, the percentage-difference sensitivity analysis technique is applied.

The percentage-difference sensitivity analysis technique is characterized by a sensitivity index known as $S_{\theta}(Y)$, which indicates the impact of a given measure known as $Y$ for a parameter $\theta$.

Equation \ref{percentage} shows how the percentage difference index is calculated for a metric $Y(\theta)$, where $max\left \{ Y(\theta) \right \}$ and $min\left \{ Y(\theta) \right \}$ are the maximum and minimum output values, respectively, computed when varying the parameter $\theta$ over the range of its \textit{n} possible values of interest. If $Y(\theta)$ is known to vary monotonically, only the extreme values of $\theta$ (i.e., $\theta_{1}$ and $\theta_{n}$) may be used to compute $max\left \{ Y(\theta) \right \}$; $min\left \{ Y(\theta) \right \}$, and subsequently $S_{\theta}$($Y(\theta)$) \cite{matos2015}.

\begin{equation}
\label{percentage}
\text{S}_{\theta}(\text{Y}) = \frac{\text{max}\left \{ \text{Y}(\theta) \right \}-\text{min}\left \{ \text{Y}(\theta) \right \}}{\text{max}\left \{\text{Y}(\theta) \right \}}
\end{equation}

Each $S_{\theta}(Y)$ is calculated by fixing the other parameters' values. The respective impact is computed for each input parameter, and then the most significant impact is found.

\subsection{Capacity-oriented Availability (COA)}
The assurance of resource availability, considering the occurrence of failures, is a significant challenge for planning a cloud computing infrastructure and services; the same applies to blockchain application. Many techniques have been employed to measure system availability based on the average capacity available - Capacity-oriented availability-(COA) \cite{melo2017capacity}~\cite{matos2017redundant}. 

The COA assesses how the system is delivered, therefore, does not consider only states of availability or unavailability, but the real impact of these factors in service delivery. The COA calculation considers $pc_{i}$ as the operational processing capacity or the number of available resources at any state $s_{i}$, while $\pi_{i}$ describes the steady-state availability for the $s_{i}$ state, a set of all available states is known as \textit{US} state, also the maximum processing capacity of the system can be represented by \textit{MPC}. Thus, we can calculate the capacity-oriented availability by solving Equation \ref{eq:coa} \cite{paulo2011}: 

\begin{equation}
\label{eq:coa}
\text{COA} = \frac{\sum_{s_{i \in US} }pc_{i} \times \pi_{i}}{\text{MPC}}
\end{equation}

\section{Methodology}
\label{sec:methodology}
This section presents how we had accomplished our main goals and how this work can be replicated. At first, the required hardware and software resources are defined, as well as their expected behavior. The list of requirements apply to any computer system, including cloud platforms and their respective services, and corresponds to one of the steps of the system's understanding process \cite{jain1991}.

The prerequisites to create a blockchain environment based on the Ethereum private platform image include any modern hardware running Linux, MacOSX, or Microsoft Windows 10 Operating System with support to Docker Engine 18+. The Docker Engine is responsible for creating and managing the Ethereum containers.

\begin{figure}[htpb]
    \centering
    \includegraphics[width=.95\textwidth]{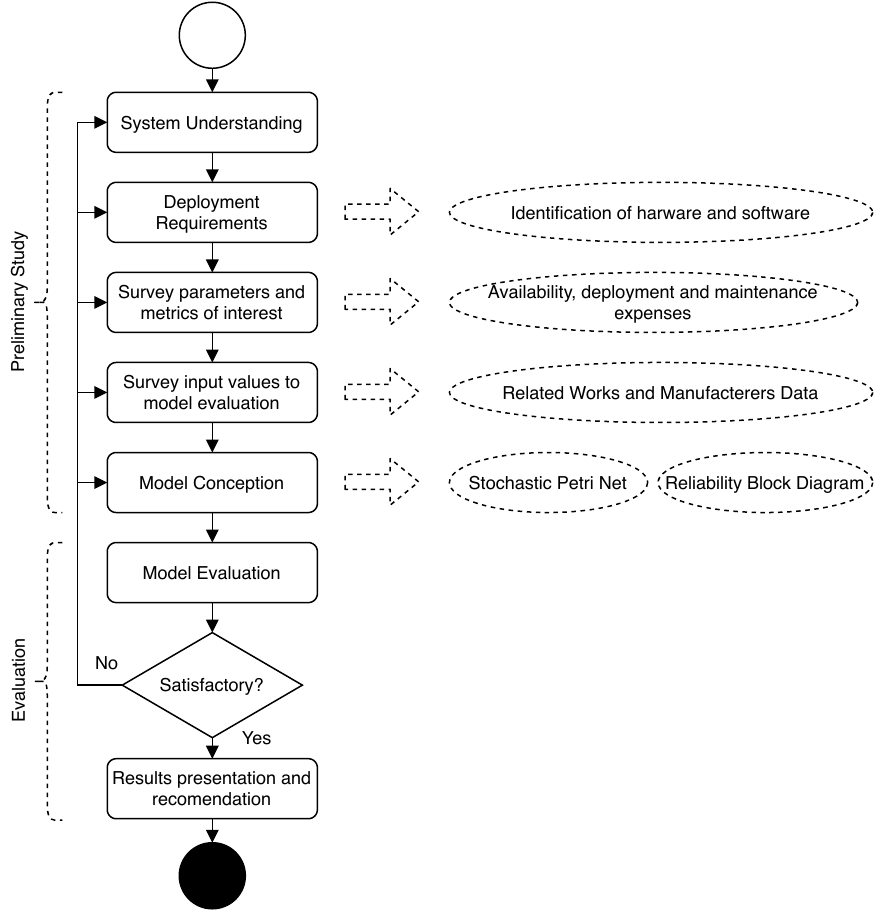}
    \caption{Supporting Methodology}
    \label{fig:ochart}
\end{figure}

After listing our relevant requirements and understand how they interact with each other, we may determine a way to evaluate the system, which includes a path to obtain the needed information about it \cite{jain1991}. The Figure \ref{fig:ochart} shows an organization chart that summarizes our strategy.

The rectangles represent each step of the methodology, and arrows connecting the rectangles define the execution order. The evaluator only advances to the next step after the completion of the current one. Already the diamond represents a step that can lead to two different paths, that is if the results are considered satisfactory, the evaluation proceeds, otherwise, it returns to any of the previous steps, where adjustments will be made until it shows a behavior close to the one expected from a real system. The dashed circles distinguish the subcategories of the methodology steps.
\subsection{Preliminary Study}
The preliminary study covers the first five steps of the supporting methodology: (1) system understanding; (2) deployment requirements; (3) survey of the parameters and metrics that will be evaluated; (4) a survey of the input values for the models, which are proposed in (5).

\textbf{System understanding}: this step consists of understanding the Ethereum platform as well as the relation between the Docker container engine and the Ethereum image.

\textbf{Deployment Requirements}: Identification of the main requirements for service deployment.

\textbf{Survey parameters and metrics of interest}: The third step corresponds to the establishment of the system metrics that will be evaluated. Choosing metrics and evaluation parameters that have a low impact on the user and administrator perception will lead to a waste of time and resources \cite{jain1991}.

\textbf{Survey input values to models evaluation}: Based on the system understanding and as a result of previous activities, we may obtain the input values that will be used by the proposed models by consulting previous knowledge, related works, and manufacturers´ specifications.

\textbf{Models Conception}: after defining the parameters, metrics of interest, and the main components required to run the service, it is possible to propose high-level models that allow obtaining a set of preliminary results.

\subsection{Evaluation}
\label{sec:evaluation}
The evaluation process has two main steps: (I) model evaluation and (II) the presentation and recommendations based on the obtained results. 

\textbf{Model Evaluation}: the proposed models were evaluated through the Mercury tool \cite{silva2017}, right after it be feeded with an input of the obtained mean time to failure (MTTF) and mean time to repair (MTTR) values; as an output, this step provides the required artifacts to accomplish the data presentation process.

\textbf{Results presentation and recommendation} This step is characterized by data representation, which can be done through graphs and tables and will depend on the audience to which the analyst wishes to present the results. The results will include the proposed models, the availability evaluation of these models, a sensitivity analysis evaluation to identify the components with the highest impact over the availability, and capacity-oriented availability evaluation. The models, evaluations, and raw data can be seen and understood by designers, analysts, and administrators, but hardly by top management; hence it important to provide graphical and textual interpretation available to support the decision-making process. 
\section{Proposed Architecture and Availability Model}
\label{sec:architectures}
\begin{figure}[htpb]
    \centering
    \includegraphics[width=.7\textwidth]{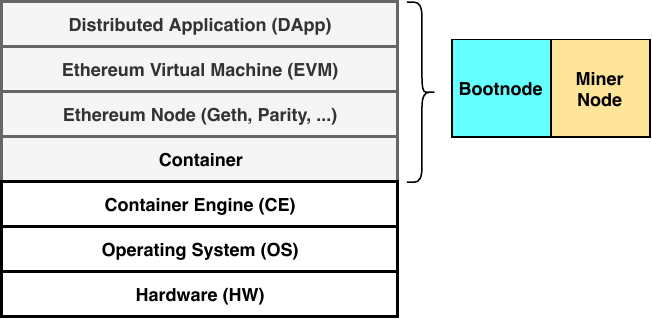}
    \caption{Service Stack}
    \label{fig:environment_stack}
\end{figure}

This section presents the evaluated environment, considering a baseline architecture managed by a single organization, with a single machine, this environment can host several Ethereum containers or nodes, that could handle or perform transactions for one or more distributed applications. Figure \ref{fig:environment_stack} shows a high-level view of the proposed environment. This Figure is based on Ethereum original paper \cite{buterin2013ethereum} and on cloud service providers, such as Amazon and Google.

As can be seen in Figure \ref{fig:environment_stack}, there are four different components in the proposed environment: the server's Hardware (HW), Operating System (OS), container engine (Docker Engine), and the deployed nodes. 

The hardware on this stack considers only the motherboard, CPU, RAM, network, disk, and so on. A power supply unit (PSU) is not considered because the meantime to failure of a component of this type is too high and the impact over the general availability is too low when compared with the other hardware components.

Already an Ethereum node can be both client and server, but it is important to mention that in this paper, we model, evaluate, and consider only the service provider side. That means that the client-side, which could include a client application, or even the network service and the respective Internet Service Provider (ISP) will not be considered for simplification.

The boot node is a dummy node, which can be used only to connect the pairs nodes or other containers so that they can exchange data, transactions, and blocks between themselves, the boot node does not execute any job that requires considerable computational power. However, it is essential because when an application needs to start or repair, a miner node needs to connect to the boot node first. The miner node is responsible execute computationally intensive jobs by performing transactions and solving puzzles to obtain the credentials to write on the Blockchain.

\begin{figure}[htpb]
    \centering
    \includegraphics[width=.7\textwidth]{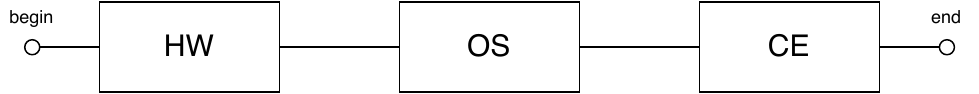}
    \caption{Server's RBD}
    \label{fig:rbd}
\end{figure}

As already seen in Figure \ref{fig:environment_stack}, all of the required components for service provisioning are hosted into a single machine. To represent this architecture and the fact that all components must be operational, we considered two-stage modeling. In the first stage, RBDs are adopted to represent nodes. The RBD presented in Figure \ref{fig:rbd} depicts the node´s components.

\begin{figure}[htpb]
    \centering
    \includegraphics[width=.7\textwidth]{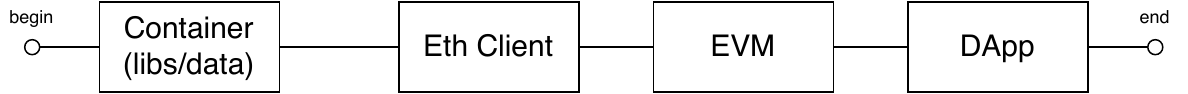}
    \caption{Containers' RBD}
    \label{fig:evm_rbd}
\end{figure}

 This RBD´s components are in series since if one component fails, the whole subsystem fails. After evaluating the server RBD, the respective MTTF is obtained. The MTTR values are discussed later on. Then, we have modeled both nodes, the miner, and the boot node through another RBD. Figure \ref{fig:evm_rbd} presents the containers RBD, which can be used to represent both types of Ethereum nodes.
 
\begin{figure}[htpb]
    \centering
    \includegraphics[width=.7\textwidth]{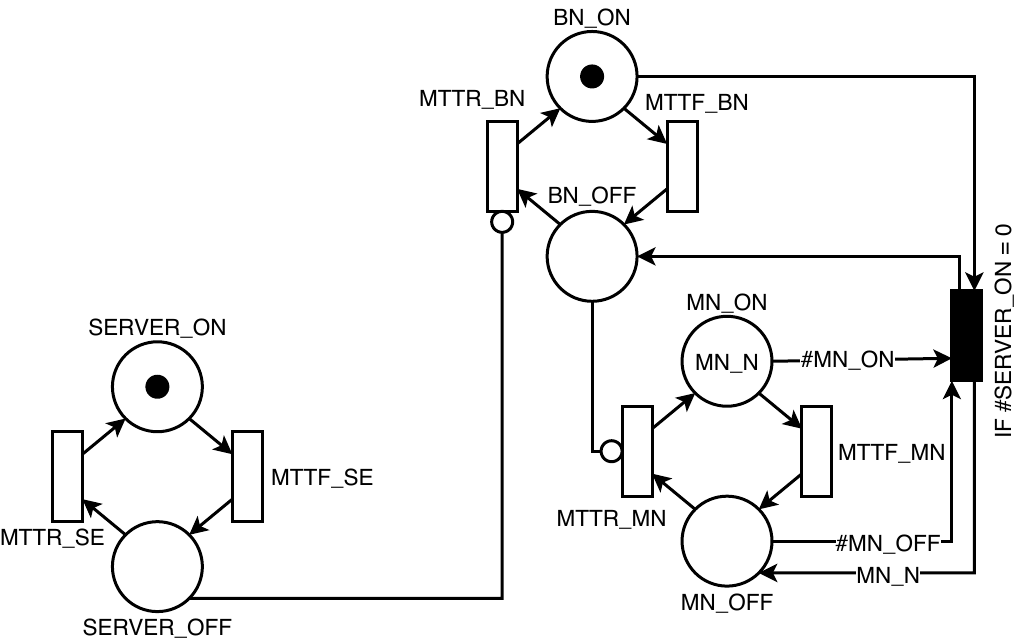}
    \caption{Service Provisioning - SPN Model}
    \label{fig:baseline}
\end{figure}

Finally, Figure \ref{fig:baseline} presents the top-level environment model, which hierarchically relates the two RBDs previously depicted and representing each of the main components in the Ethereum environment. The Service Infrastructure is represented by a Stochastic Petri Net (SPN) due to the dependency between the miner, boot node, and the container engine which is a part of the server.

This SPN depicts three components. The first component is the basic server, whose MTTF\_SE and MTTR\_SE are obtained through the evaluation of the previously presented RBD (See Figure \ref{fig:rbd}).  The other two components seem in Figure \ref{fig:environment_stack} are represented (each) by a single component RBD (See Figure \ref{fig:evm_rbd}). The places (circles) Server\_ON, BN\_ON, and MN\_ON mark the upstate for each component, respectively Server, Bootnode, and Miner Node containers, while their counterpart, that means a mark on places Server\_OFF, BN\_OFF, and MN\_OFF indicates that the component is in a downstate. 

Initially, every component is operational or in upstate, each component has a mean time to fail (MTTF) and a mean time to repair (MTTR), these times follow an exponential distribution, and are represented in the model by timed transitions (white squares) connected through arcs (arrows) to the place that corresponds to the component current state (ON or OFF, UP or DOWN).

There are two types of arcs in the proposed model, the arc that ends with an arrow is the most common, and indicates the activation of a transition if the requirements are met and the movement of a token (the small black circle inside a place, which indicates the number of resources), while the arc that ends with a circle is called inhibitor arc, there are four of them in the proposed model. 
An inhibitor arc indicates that a transition can only be activated if some conditions are met, for example, if there is no boot node operational and we need to repair a miner node, the miner node can only be repaired if the boot node has been repaired first, which can be seen as a dependency relationship.

Finally, there are two types of transitions, the already mentioned timed transition that depends on time to activate and the black rectangle, which is the transition called immediate transition, which may be activated if the required amount of tokens are present in the connected place. The proposed model has one immediate transition, used to change the state of a component if this component is dependent from another that failed, for example, if the server enters into a failure state all nodes (miners and boot node) will fail as well, meaning that in this case, the system does not wait for the mean time to fail associated to the timed transitions.
\section{Case Studies}
\label{sec:case}
This section provides three case studies that demonstrate how feasible are the proposed models: (1) Availability evaluation, (2) Capacity-oriented Availability (COA) evaluation, both of them regarding an application or service provisioning, (3) cost evaluation and comparison between public and private service providers. The first step to accomplish our tasks is to obtain the required system's input values to feed the model and perform the availability evaluation. Some of these values were obtained from a literature review \cite{dantas2013,melo2016,sebastio2018availability}, while other ones were extracted from manufacturer charts and white papers. The values used in this paper can be seen in Table \ref{tab:input_baseline}. It is important to mention that software-based component has an estimated mean time to repair and can be obtained by measuring the time that usually takes to create or start a new service or component, that tends to be a lower value when compared to hardware components that may demand an entirely new motherboard, CPU or disk to be changed. 

\begin{table}[htpb]
\centering
\footnotesize
\caption{Input Parameters for Proposed Models}
\label{tab:input_baseline}

\begin{tabular}{l|r|r}
\hline
\multicolumn{1}{c|}{Component}   & \multicolumn{1}{c|}{MTTF (h)} & \multicolumn{1}{c}{MTTR (h)} \\ \hline
Hardware (HW)  & 8760 & 1.66 \\
Operating System (OS) & 2893 & 0.25 \\
Docker Engine (DE) & 2516 & 0.25 \\
Container & 1258 & 0.25 \\
EVM & 788.4 & 0.25 \\
Eth Client & 788.4 & 0.25 \\
DApp & 788.4 & 0.25 \\ \hline
\end{tabular}
\end{table}-

\subsection{Case Study I - Availability Evaluation}
\label{subsec:case1}
The first case study is the availability evaluation of the proposed models. It is important to mention that this evaluation was performed using the Mercury Tool \cite{silva2017}. 

We first have evaluated the RBD presented in Figure \ref{fig:rbd}, which could be represented by the expression denoted by $\text{A}_{\text{HW}} \times \text{A}_{\text{OS}} \times \text{A}_{\text{CE}}$, where A stands for Availability. The obtained availability value for this model was 99.96\%, corresponding to an annual downtime of 3.28 hours, while the obtained MTTF 1166.48 hours and the respective MTTR was 0.4378 hours. 

The second evaluated model was the RBD for the Ethereum nodes (See Figure \ref{fig:evm_rbd}), this evaluation apply to both miner and bootnode, the closed-form equation that represent this model is $\text{A}_{\text{Container}} \times \text{A}_{\text{Eth Client}} \times \text{A}_{\text{EVM}}  \times \text{A}_{\text{DApp}}$, the obtained values were 99.9081\% of availability, for a MTTF of 271.9 hours and a MTTR of 0.25 hours. 

The values obtained from the RBDs evaluation were later used as an input parameter for the Service Infrastructure model, which is the SPN presented in Figure \ref{fig:baseline}. The general availability obtained by the SPN evaluation is 99.82\%, resulting in an annual downtime of 8.78 hours. During the downtime period, users can not request or send new transactions to the system. Also, we must mention that only a single miner node was operational on the considered evaluation, we could have many others, as we are going to see on the COA case study. This model's availability was given by the probability of having at least one token in the MN\_ON place since, as already mentioned, the boot node is only required when we need to start or repair a miner node. Table \ref{tab:general} resumes the availability metrics obtained through model evaluation.

\begin{table}[htpb]
\footnotesize
\centering
\caption{General Availability Values}
\begin{tabular}{l|r|r|r|r}
\hline
\multicolumn{1}{c|}{\textbf{Model}} & \multicolumn{1}{c|}{\textbf{Availability}} & \multicolumn{1}{l|}{\textbf{Downtime(h/y)}} & \multicolumn{1}{c|}{\textbf{MTTF(h)}} & \multicolumn{1}{c}{\textbf{MTTR(h)}} \\ \hline
Server                       & 99.96                                      & 3.28                                              & 1166.48                                & 0.43                                  \\
Containers                     & 99.90                                      & 8.76                                              & 271.9                                  & 0.25                                  \\ \hline
SPN                    & 99.82                                      & 15.77                                             & 990                                     & 1.7                                  \\ \hline
\end{tabular}
\label{tab:general}
\end{table}

As an attempt to detect the components with the highest impact over the system availability, we have applied a sensitivity analysis Through the percentage difference technique using the Mercury tool. To accomplish this task, we fed the model with the required MTTF, and MTTR values for each component, the values presented in Table \ref{tab:input_baseline} will be reused. The adopted values vary from +50\% to -50\% from the original value. Table \ref{tab:output_baseline} presents the sensitivity analysis results obtained from the baseline architecture availability evaluation. It is important to mention that the sensitivity analysis can also be applied to check if the used MTTF and MTTR values are close enough to what can be obtained in a real system and to point out where improvements can be made. 

\begin{table}[htpb]
\centering
\begin{tabular}{l|r|r}
\hline
& \multicolumn{2}{c}{Percentage Difference Sensitivity Analysis}\\
\hline
\multicolumn{1}{c|}{Parameters} & \multicolumn{1}{c|}{Ranking} & \multicolumn{1}{c}{Sensitivity Index} \\ \hline
$\text{MTTF}_{\text{SE}}$ & 1st & $0.00109$ \\
$\text{MTTR}_{\text{MN}}$ & 2nd & $0.00100$\\
$\text{MTTR}_{\text{SE}}$ & 3rd & $9.37 \times 10^{-4}$\\
$\text{\#Miners}$ & 4th & $6.26 \times 10^{-4}$\\
$\text{MTTR}_{\text{SE}}$ & 5th & $4.86 \times 10^{-4}$\\
$\text{MTTR}_{\text{BN}}$ & 6th & $1.91 \times 10^{-4}$\\
$\text{MTTF}_{\text{BN}}$ & 7th & $1.32 \times 10^{-6}$\\
\hline
\end{tabular}
\caption{Sensitivity Analysis Ranking}
\label{tab:output_baseline}
\end{table}

\begin{figure*}[htpb]
    \centering
    \subfloat[][]{\label{fig:mttf_server}\includegraphics[scale=0.5]{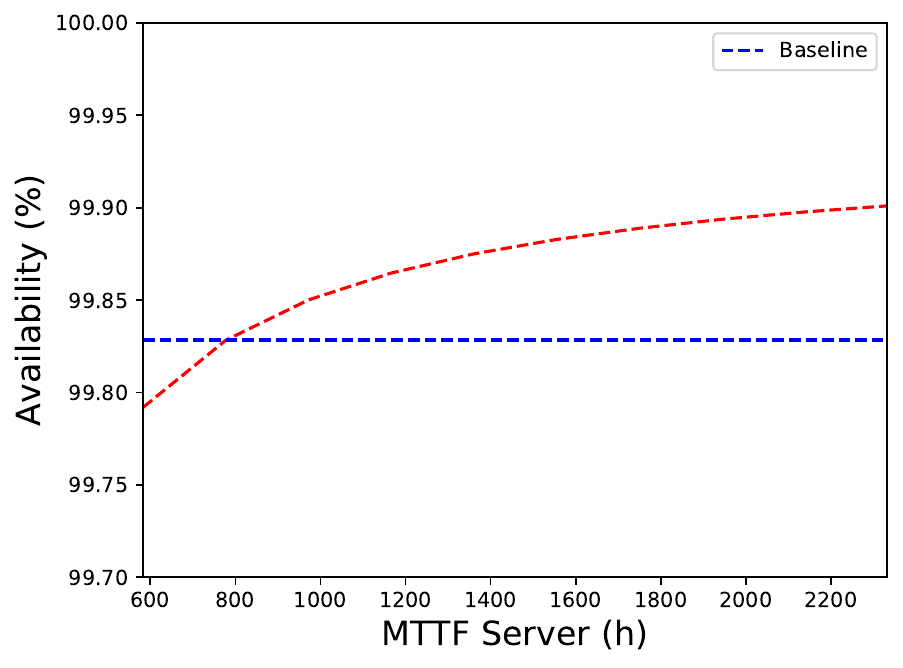}}
    \subfloat[][]{\label{fig:mttr_mn}\includegraphics[scale=0.5]{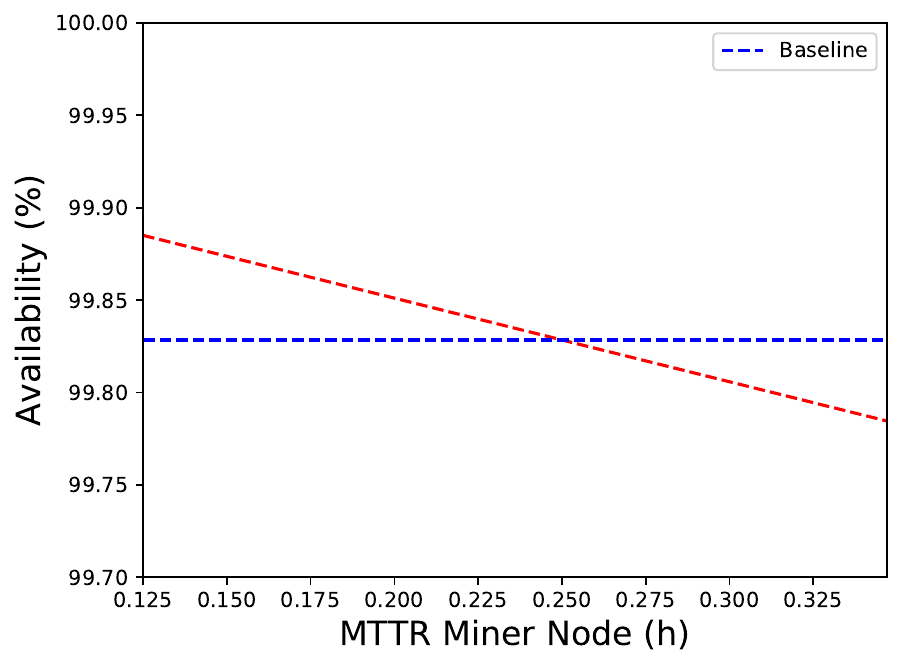}}
    \qquad
    \subfloat[][]{\label{fig:mttr_server}\includegraphics[scale=0.5]{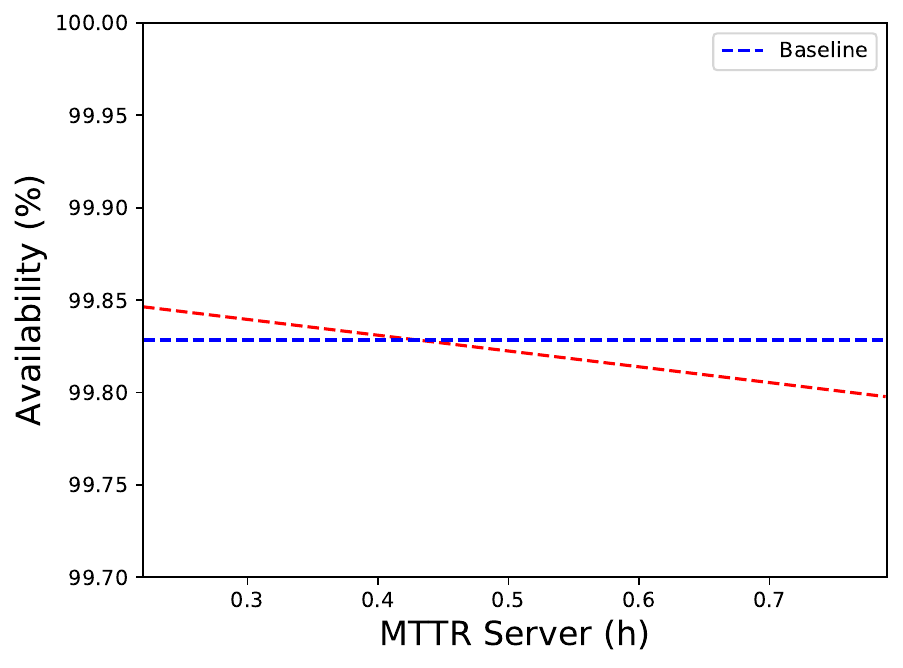}}
    \subfloat[][]{\label{fig:mttf_mn}\includegraphics[scale=0.5]{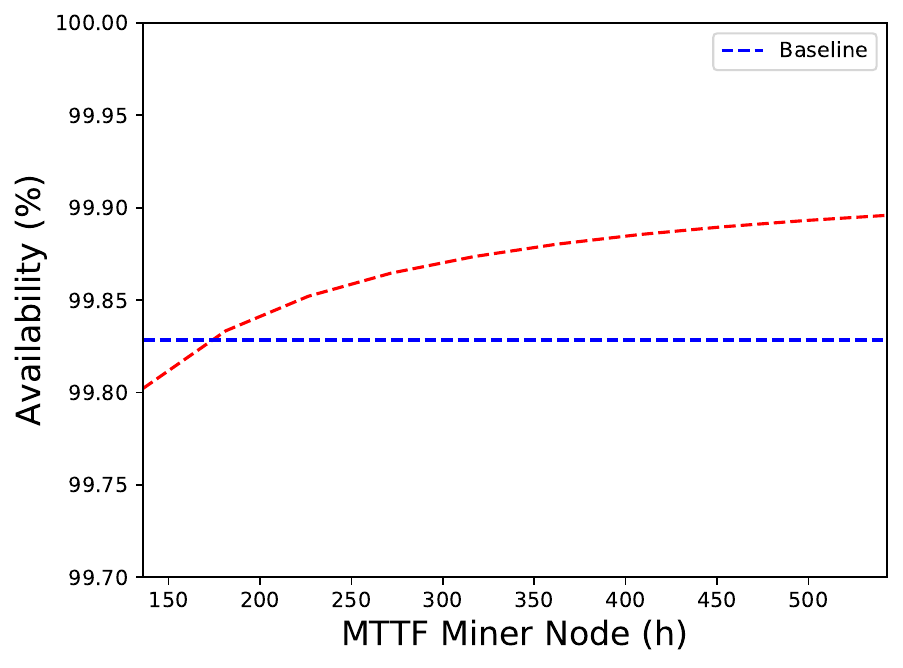}}
    \qquad
    \subfloat[][]{\label{fig:mttr_bn}\includegraphics[scale=0.5]{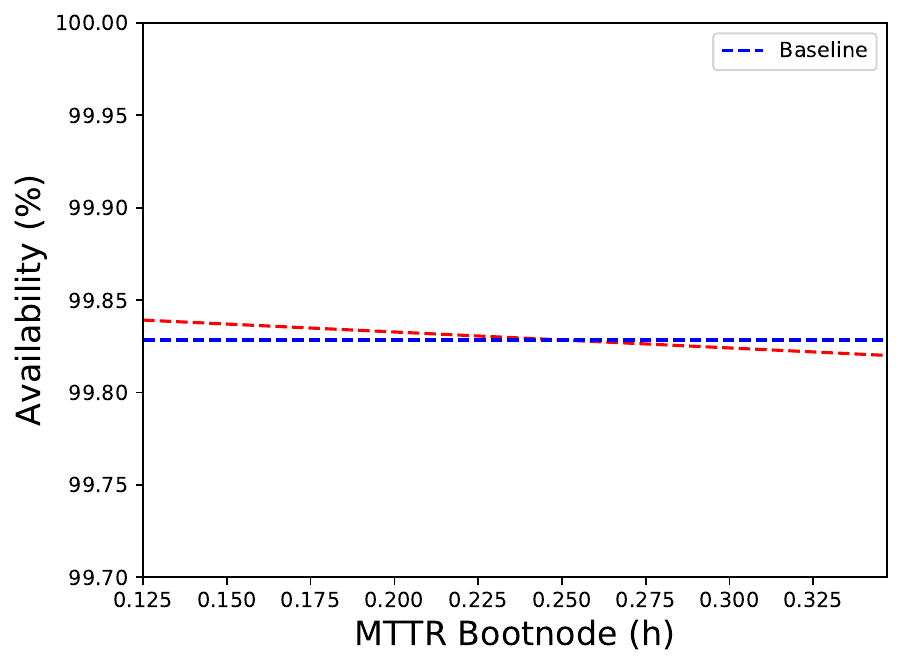}}
    \subfloat[][]{\label{fig:mttf_bn}\includegraphics[scale=0.5]{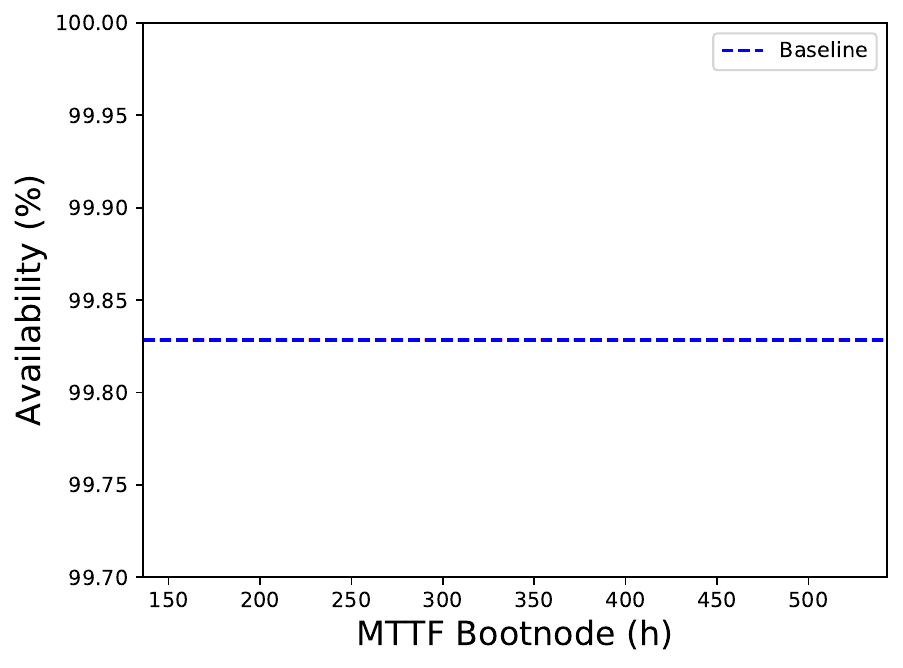}}
    \caption{Sensitivity Analysis}
    \label{fig:sensitivity}
\end{figure*}

The presented indexes start from the factor with the highest impact over the architecture availability is the MTTF of the server (which is the component that is represented by the RBD of Figure \ref{fig:rbd} and includes hardware, operating system, and contained engine), which is followed by the MN MTTR and Server MTTR. These components had a higher impact on the environment availability probably because their failure directly impacts the failure of other components; every single component will fail if the server enters into a failure state. Figure \ref{fig:sensitivity} shows the behavior of all the factors over the baseline infrastructure availability based on the passage of time in the Percentage-Difference sensitivity analysis evaluation.

The Figure \ref{fig:sensitivity} presents a graph for each component and their respective MTTR and MTTF impact over the general availability, the red line on each graph represents the variability of the system's availability based on MTTF and MTTR values changes, has the MTTF of the boot node in the last position. This component presents a minor impact over the system's availability since it can enter into a failure state, and the service will be provided as long as there is a miner node running.

\begin{figure}[htpb]
    \centering
    \includegraphics[width=.7\textwidth]{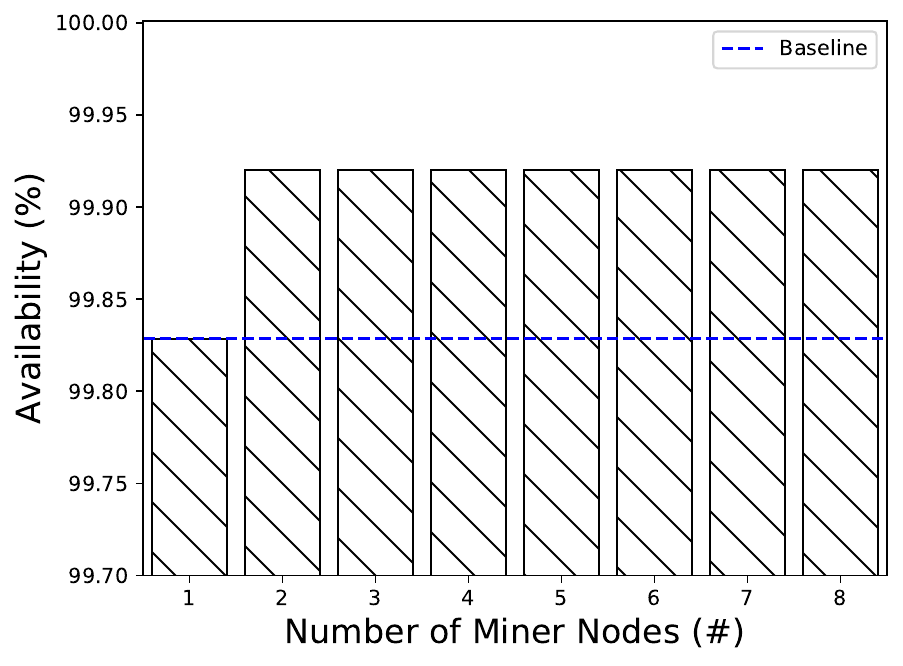}
    \caption{Impact of the number of miner nodes over the availability}
    \label{fig:sensitivity_mn}
\end{figure}

Another essential component is the number of miner nodes. Here, we varied it from 1 to 8 miner nodes and checked the impact of this component on system availability. Figure \ref{fig:sensitivity_mn} presents the impact of this component on system availability. From 1 to 2 miner nodes there is a real gain on availability, but after this, two or more, the improvement is almost none.

\begin{figure}[htpb]
    \centering
    \includegraphics[width=.7\textwidth]{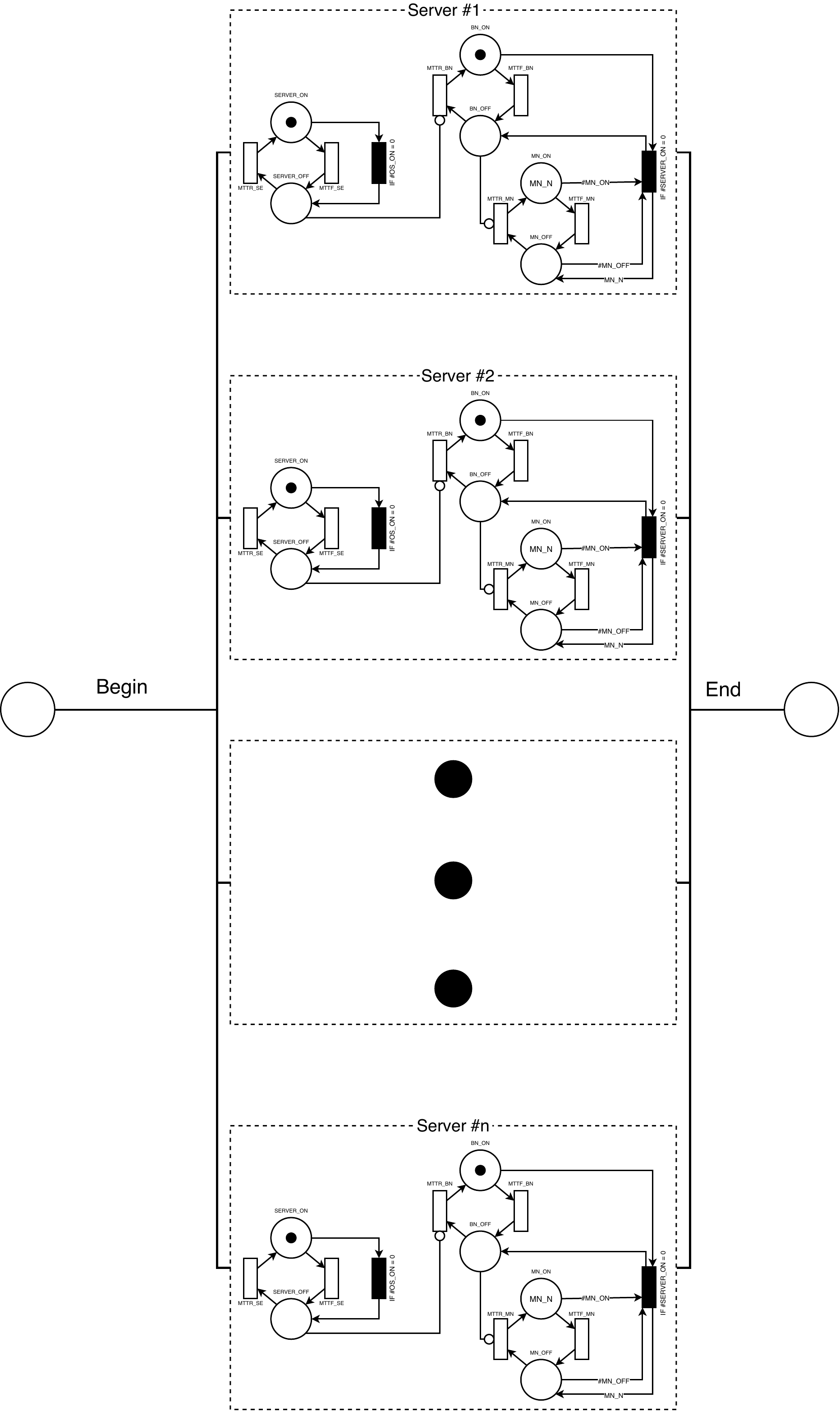}
    \caption{Multi-server environment}
    \label{fig:redundante}
\end{figure}

As an attempt to improve the system's availability, we provide a scenario considering multiple nodes in a parallel way, aiming to reach high availability. Using models in a hierarchical way we could distribute the servers and service provisioning in a parallel way, by using a reliability block diagram, we reach to a scenario that is closer to a real-world environment, which is presented in Figure \ref{fig:redundante}.

\begin{figure}[H]
    \centering
    \includegraphics[width=.7\textwidth]{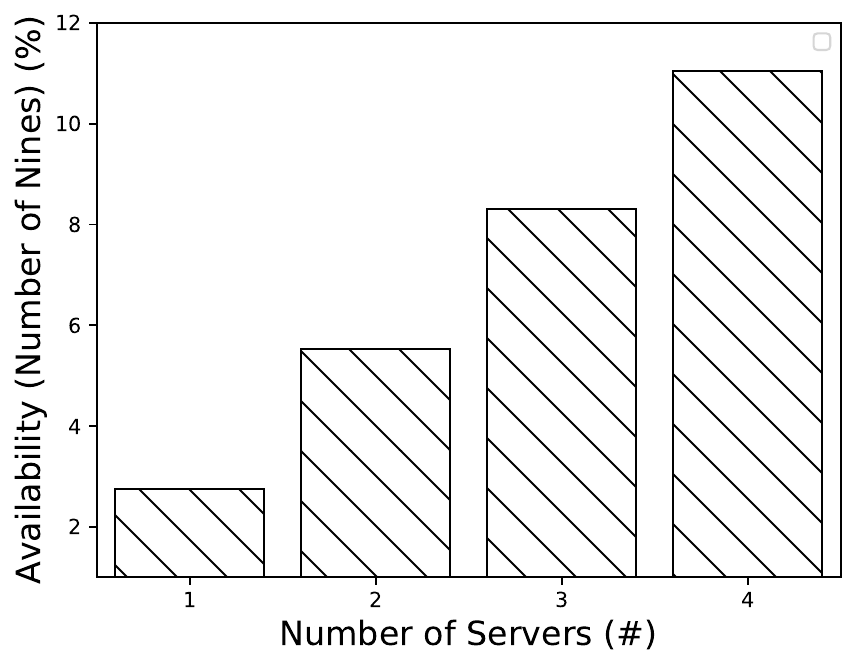}
    \caption{Availability values for multiple nodes}
    \label{fig:av4}
\end{figure}

As an example, we considered up to 4 servers and presented the availability evolution in Figure \ref{fig:av4}. The general availability reaches five-nines when considering just two nodes (99.999\%), implying in an annual downtime of fewer than five minutes.

If we keep adding servers the availability value will have almost no impact on service provisioning, but it will improve the general expenses and on general performance, and in the case of most Blockchain-based applications, the larger the amount of nodes the higher is the time taken to validate or perform a transaction. 
\subsection{Case Study II - Capacity-oriented Availability}
\label{subsec:case2}
The Capacity-oriented Availability (COA) may be estimated with Markov Reward Models (MRMs), by assigning a reward to each model state corresponding to the system capacity in that condition. Another way of estimating COA is by computing probabilities of being at each system state, multiplying them by the capacity associated with each state, summing up these products, and dividing this result by the maximal system processing capacity. In a private cloud system, COA may be associated with the number of virtual machines or containers that the servers can instantiate, therefore representing the computational work the system can provide regarding the availability values previously obtained.

The capacity-oriented availability of the cloud infrastructure is calculated by dividing the average of service capacity the system can deliver for the maximal service capacity. Therefore, the COA for any configuration of cloud infrastructure can be found following the closed-form equation adapted from ~\cite{dantas2019}, which can be expressed by Equation \ref{eq:coaall}.

\footnotesize
\begin{multline}
\label{eq:coaall}
    COA=\frac{1}{n}((\sum_{k=0}^n ( (A_{s} )^p CTs[k,n](n-k)) )
    + \\ ( \sum_{j=1}^{p-1}\sum_{k=0}^{NCTj}(A_{s}(j,p)-A_{s}(j+1,p)) \times (CTs[k,NCTj])(NCTj-k) )),
\end{multline}
\normalsize

\noindent where:

\footnotesize
\begin{equation}
\label{eq:vms}
    CTs[k\_,n\_]=\left (\frac{\frac{n!}{(n-k)!}\mu^{(n-k)}\lambda_{ct}^k)}{\sum_{i=0}^n \frac{n!}{i!}\lambda_{ct}^{n-i}\mu^i} \right )
\end{equation}
\normalsize

\begin{itemize}
    \item $CTs[k\_,n\_]$ is a function which defines the probability of state $k$, from $n$ containers available $(P_{k})$;
    
    \item $NCTj$ represents the number of containers $(NCT)$ per physical machine $(j)$.
\end{itemize}

\begin{figure}[htpb]
    \centering
    \includegraphics[width=.7\textwidth]{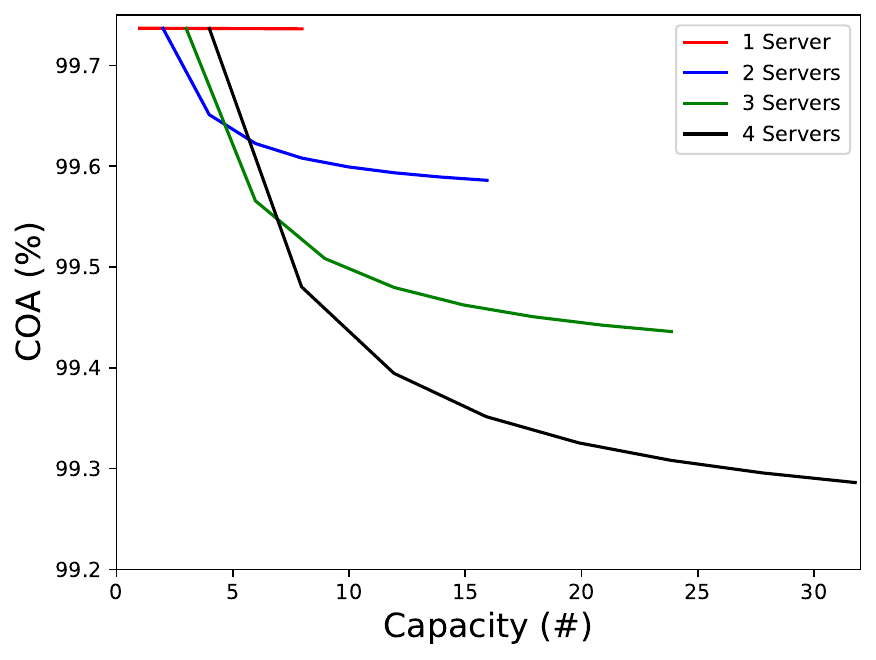}
    \caption{COA x System's Capacity}
    \label{fig:coa}
\end{figure}

For the proposed case study, a single server can handle a container per core, which means that if the server CPU has eight cores, then it can run up to 8 miner nodes, each one running the Ethereum Virtual Machine and a personal copy of the blockchain/service in its current state. As previously mentioned, the boot node is a dumb node, and it is only required when we need to start a new miner node or repair one that entered into a failure state. So, as expected, we do not consider its impact on the environment's capacity-oriented availability (COA). We have measured the COA value by varying the number of available resources from 1 to 8, considering an octa-core processor, Figure \ref{fig:coa} presents the capacity for a scenario with 1, 2, 3, and 4 servers, each one running up to 8 containers.

As shown in Figure \ref{fig:coa}, we may see that the more significant the amount of available resource, the lower the Capacity-oriented Availability, which happens since with more resources, we have a higher probability that at least one of these resources enter into a failure state.

\subsection{Case Study III - Cost Planning}
\label{costs}

A three-stage survey characterizes the cost analysis. Which has as a prerequisite the COA analysis previously performed. At the first stage of the survey, the necessary components to provide a cloud computing environment were defined, such as servers, switch, rack, and air conditioning.

In the second stage of the survey, all the acquisition costs for each component were individually analyzed, and by consulting Brazilian and North American websites, the lowest value of each architecture element in the period from 8th to May 14th, 2020, was selected. Table \ref{tab:aqcosts} present the cost for each component, as well as the store with the lowest price.

\begin{table}[H]
\footnotesize
\centering
\caption{Acquisition Costs}
\begin{tabular}{l|r|r}
\hline
\multicolumn{1}{c|}{\textbf{Component}} & \multicolumn{1}{c|}{\textbf{Cost (USD)}} & \multicolumn{1}{c}{\textbf{Store}} \\ \hline
HPE ProLiant DL360 Gen10                & 1,884                                    & Amazon                              \\
Switch Huacomm HC1709P                  & 63                                       & Amazon                              \\
Air conditionair Black+Decker BPACT08   & 310                                      & \multicolumn{1}{l}{Amazon}         \\
Rack SRW 15.900                         & 320                                      & Sysracks                            \\ \hline
\end{tabular}
\label{tab:aqcosts}
\end{table}

The third and last stage of our survey was the technical specification analysis of each environment component, in search of energy power consumption values to calculate the energy cost per unit for one year period. By using the HPE Power Advisor\footnote{https://paonline56.itcs.hpe.com/?Page=Index} to estimate the data center energy consumption, we could achieve real values for the energy consumption of the selected components by using Equation \ref{kwh}.

\begin{equation}
\label{kwh}
\footnotesize
\text{kWh} = \frac{\text{Power (W)} \times \text{NHD} \times \text{NDY}}{1000}
\end{equation}
where NHD stands for the number of hours per day and NDY is the number of days per year that the equipment is operational, so considering that each component would be running for 24 hours a day, seven days a week and a kilowatt-hour cost of 16 cents of American Dollar we could achieve the values presented in Table \ref{tab:relacao}.

\begin{table}[htpb]
\footnotesize
\centering
\caption{Energy consumption by year}
\begin{tabular}{l|r|r}
\hline
\multicolumn{1}{c|}{\textbf{Component}} & \multicolumn{1}{c|}{\textbf{Power (W)}} & \multicolumn{1}{c}{\textbf{Energy Expenses/y (USD)}} \\ \hline
HPE ProLiant DL360 Gen10                & 122.48                                  & 171.87                                              \\
Switch Huacomm HC1709P                  & 65                                      & 91.1                                                \\
Cooling                        & 187.48                                  & 262.77                                              \\ \hline
\end{tabular}
\label{tab:relacao}
\end{table}

By considering an eight-core processor for the chosen server, and by using HP Power Advisor (HPA), we estimated the general expenses for one, two, three, and four servers. It is important to highlight that the recommended cooling expenses are equivalent to the total expenses, which means that we will probably spend a dollar with the cooling system for each dollar expended in computing. Table \ref{tab:maintenance} shows the expenses for each scenario. It is important to mention that the acquisition cost for each environment considers a fixed amount of cooling system, switch and rack, the only thing that varies is the number of servers, as can be seen in the first column. In other words, the acquisition cost is defined by the sum of the values for each component on the environment: server + switch + rack + cooling device.

\begin{table}[htpb]
\footnotesize
\centering
\caption{Maintenance Expenses by Year}
\begin{tabular}{r|r|r}
\hline
\multicolumn{1}{c|}{\textbf{\#Servers}} & \multicolumn{1}{c|}{\textbf{Total Acquisition (USD)}} & \multicolumn{1}{c}{\textbf{Maintenance Cost/y (USD)}} \\ \hline
1                                       & 2,577                                               & 525.54                                                \\
2                                       & 4,461                                               & 777.78                                                \\
3                                       & 6,345                                               & 1,636.13                                              \\
4                                       & 8,229                                               & 2,837.82                                              \\ \hline
\end{tabular}
\label{tab:maintenance}
\end{table}

By knowing how much service we can provide (almost eight containers for each server (as already seen in subsection \ref{subsec:case2})), and how much costs to provide service over environments composed by one, two, three and four nodes, we may compare the service provisioning of the most popular public cloud computing environments and the proposed private infrastructures.

\subsubsection{Public vs Private Cloud}
\label{ppc}

The chosen public cloud computing platforms for comparison to the private environment modeled here are Google Cloud, Amazon AWS, and Microsoft Azure. These three platforms provide Ethereum blockchain environments for distributed application services. 

As expected, each one has its peculiarities, we chose a basic container/VM for each platform, the settings are close to the previous stated in this paper, one vCPU, and up to 4 Gib RAM. Table \ref{tab:vmprice} summarizes the selected environments.

\begin{table}[htpb]
\footnotesize
\centering
\caption{Annual Expenses by Instance in Public Clouds}
\begin{tabular}{l|l|r|r}
\hline
\multicolumn{1}{l|}{\textbf{Platform}} & \multicolumn{1}{l|}{\textbf{Instance Type}} & \multicolumn{1}{c|}{\textbf{On-demand /y (USD)}} & \textbf{Availability}(\%)\\ \hline
Google                                 & n1-standard-1                         & 291.24   & 99.9                                             \\
Amazon                                 & m3.medium                             & 420.48  & 99.9                                              \\
Microsoft                              & D1s                                   & 403,56 & 99.9\\ \hline
\end{tabular}
\label{tab:vmprice}
\end{table}

\begin{figure}[htpb]
    \centering
    \subfloat[][]{\label{fig:eight}\includegraphics[scale=0.41]{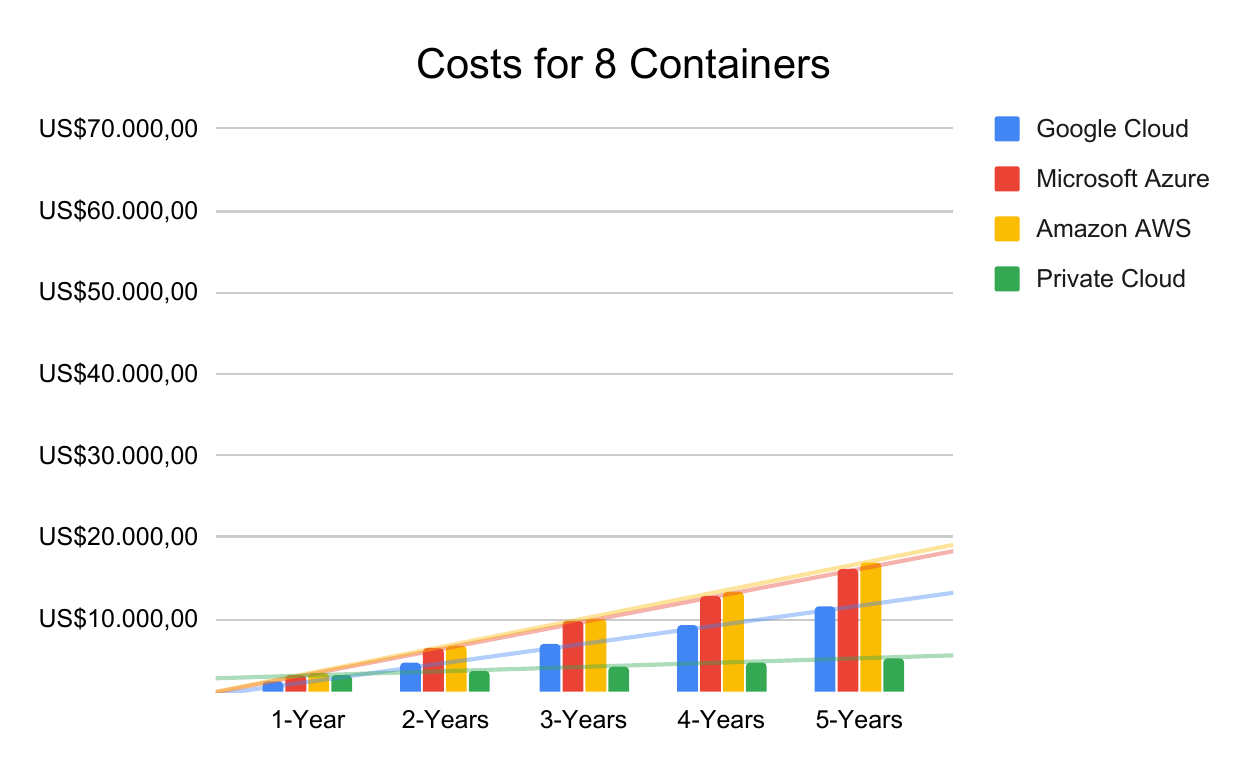}}
    \qquad
    \subfloat[][]{\label{fig:sixteen}\includegraphics[scale=0.41]{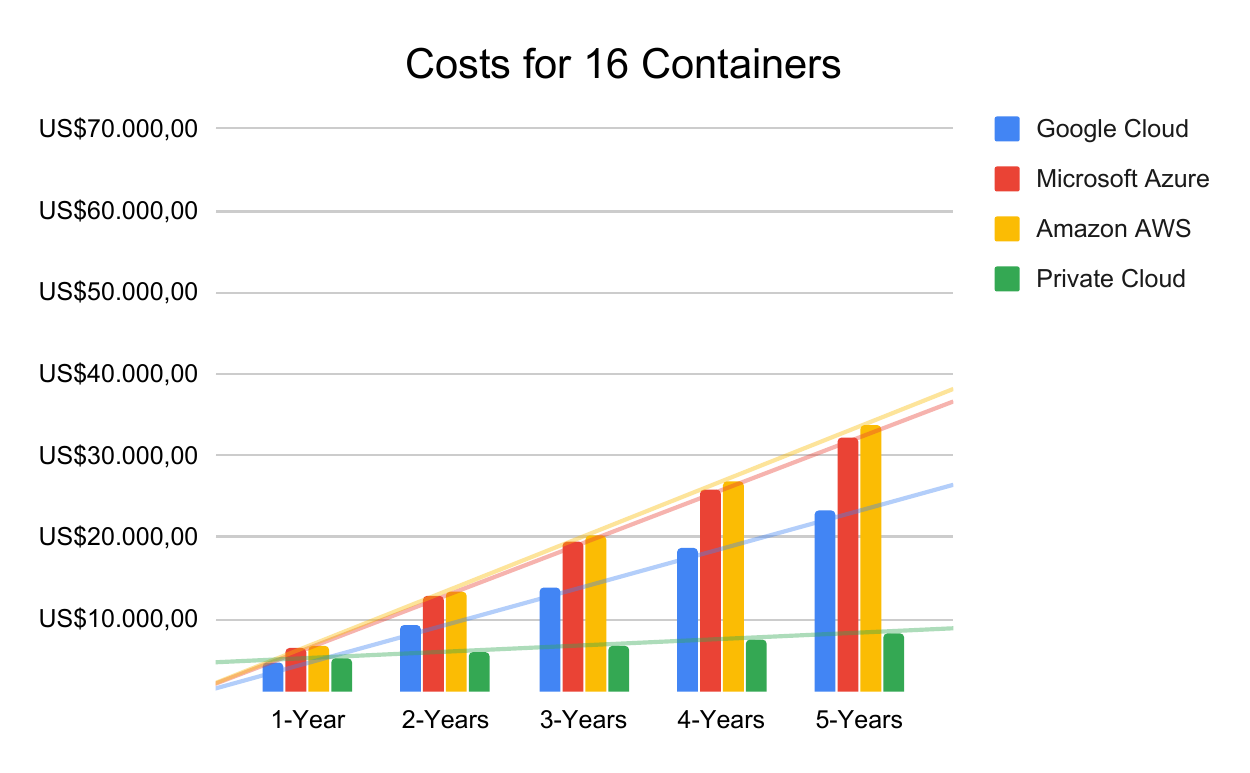}}
    \qquad
    \subfloat[][]{\label{fig:twentfour}\includegraphics[scale=0.41]{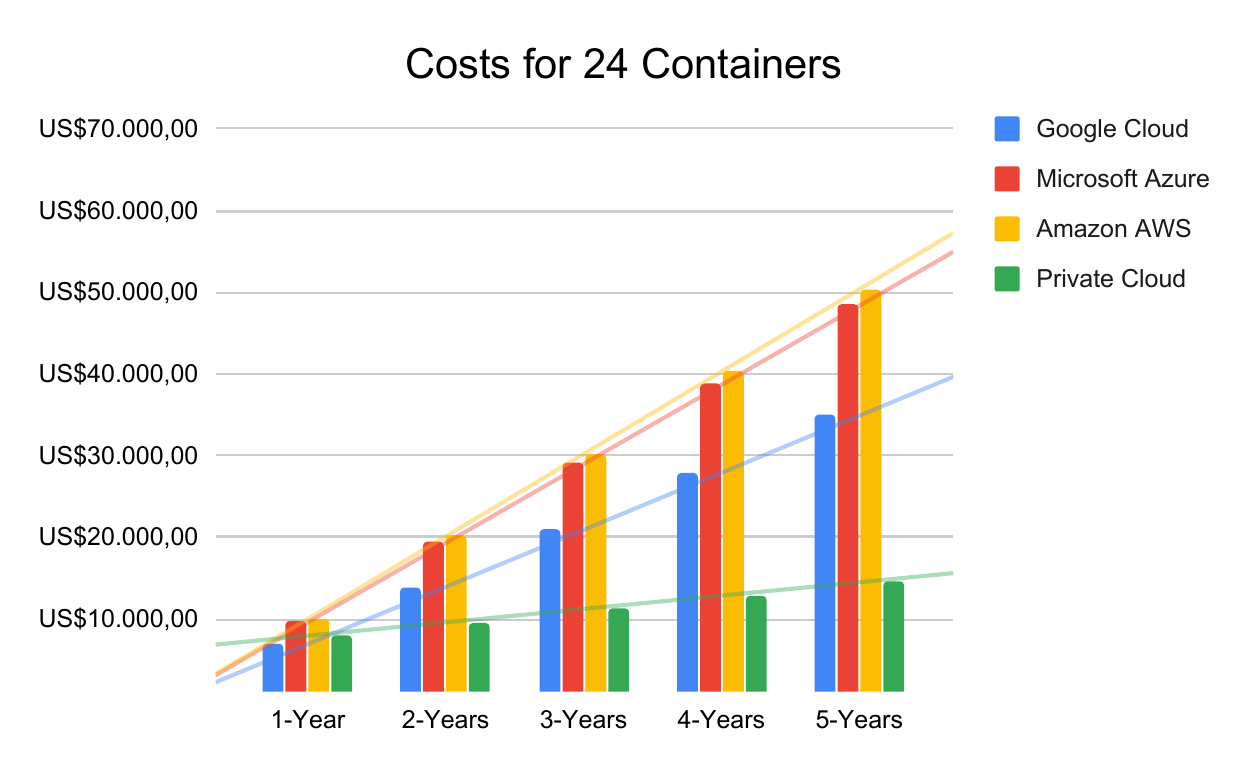}}
    \qquad
    \subfloat[][]{\label{fig:thirdtwo}\includegraphics[scale=0.41]{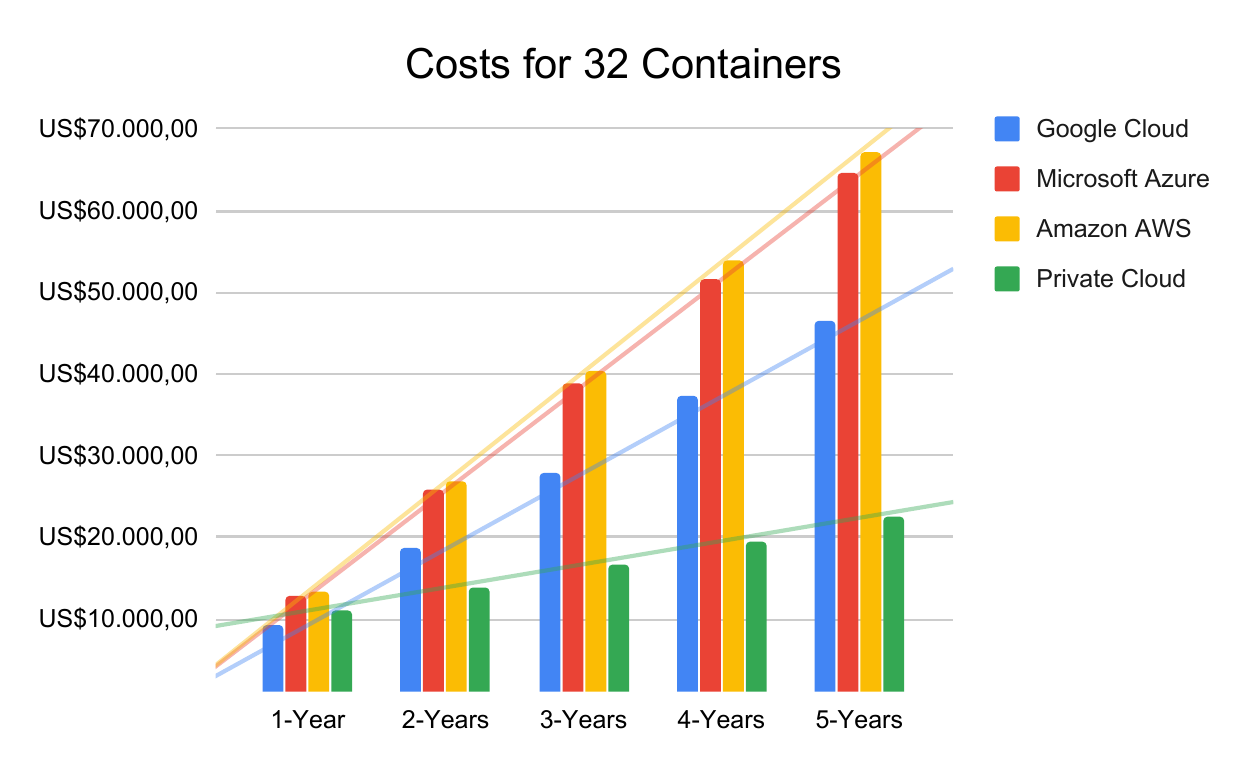}}
    \caption{Effective Expenses - Public vs Private Clouds}
    \label{fig:effetive_expenses}
\end{figure}

Finally, we compared four different scenarios considering 8, 16, 24, and 32 containers/Virtual Machines in these three cloud computing providers and the use of private cloud computing to deliver the Ethereum platform as a service. Figure \ref{fig:effetive_expenses} present the effective expenses comparison for each platform, including private and public ones.

As can be seen, from both Table \ref{tab:vmprice} and Figure \ref{fig:effetive_expenses} among the public cloud providers, the Google Cloud has the best cost. By considering the life expectancy for a server is from three to five years, we extended our analysis to accommodate this entire period. In general terms, for the first year, the proposed private environment loses to the Google Cloud, which is expected, since the equipment acquisition expenses are higher than the required maintenance expenses.

From the tendency, lines are possible to see that Amazon AWS is not a suitable choice, neither the Microsoft Azure. Finally, it is important to mention that this evaluation does not consider maintenance teams' expenses if the decision-maker has to pay someone else to do this work is probably worse to keep a private environment than to contract instances in Google cloud. However, if the distributed application requires high-security, a suitable alternative would be to adopt a private cloud infrastructure at the expense of the cost.

It is important to mention that the availability guaranteed values for those platforms in theirs respective Service Level Agreements (SLA) are 99.9\%, with up to eight hours of annual downtime, which is higher than our first evaluated scenario, that considered only one server, but lower than a scenario with at least two servers (99.999\%). 

\section{Conclusions and Future Works}
\label{sec:conclusions}

This paper proposes a set of models to evaluate the availability of a cloud computing infrastructure that provides blockchain-as-a-service. Three case studies demonstrate the feasibility of the proposed models, and help those interested in planning and use blockchains applications based on the Ethereum platform. 

The first case study dealt with the proposed models' availability evaluation through the Mercury modeling tool; the baseline environment availability hit 99.82\%, then we have performed a sensitivity analysis evaluation and identified the most critical component to provide the service was the service itself. After that, we have evaluated an environment with multiple servers through a reliability block diagram (RBD) that hierarchically distributed the SPN model in order to improve the number of resources that can be delivered to the system's users.

Already the second case study, which was a capacity-oriented availability (COA) evaluation, presented the real amount of resources that could be delivered, due to the presence of the system's failures and repairs routines. This case study showed that the more significant the amount of servers to service provisioning, the lower is the COA, which goes on the counter-hand of the service availability.

The third and last case study evaluated the expenses to provide Blockchain as a service in a private cloud environment, as well as a public platform to host a distributed application based on the Ethereum blockchain platform. Google Cloud has the lowest cost among the public cloud computing service providers, and Amazon has the worst price. The private environment presented the general lower expenses, except for the first year, considering the necessity to buy all required components to accomplish service provisioning.

As future work, we intend to evaluate cost-benefit, as well as the presence of software aging on the environment, aiming to provide a maintenance routine to help to improve the availability of a single server environment and extend the system's life cycle. 


\begin{acknowledgements}
The authors would like to thank the Brazilian Government for the financial support through the Fundação de Amparo a Ciência e Tecnologia de Pernambuco (FACEPE), the Modeling of Distributed and Concurrent Systems (MoDCS) group for the help on improving this research.
\end{acknowledgements}

\section*{Conflict of interest}

The authors declare that they have no conflict of interest.


\bibliographystyle{spmpsci}
\bibliography{sample}

\end{document}